# American Family Cohort, a data resource description

Deepa Balraj[1], Ayin Vala[2], Shiying Hao[2], Melanie Philofsky[3], Anna Tsvetkova[3], Elena Trach[3], Shravani Priya Narra[3], Oleg Zhuk[3], Mary Shamkhorskaya[3], Jim Singer[3], Joseph Mesterhazy[1], Somalee Datta[1], Isabella Chu[2], David Rehkopf[2]

1. Research Technology, Technology and Digital Solutions, Stanford Health Care and Stanford School of Medicine
2. Stanford Center for Population Health Sciences, Stanford School of Medicine
3. Odysseus Data Services Inc

## Abstract:

This manuscript is a research resource description and presents a large and novel Electronic Health Records (EHR) data resource, American Family Cohort (AFC). The AFC data is derived from Centers for Medicare and Medicaid Services (CMS) certified American Board of Family Medicine (ABFM) PRIME registry. The PRIME registry is the largest national Qualified Clinical Data Registry (QCDR) for Primary Care. The data is converted to a popular common data model, the Observational Health Data Sciences and Informatics (OHDSI) Observational Medical Outcomes Partnership (OMOP) Common Data Model (CDM).

The resource presents approximately 90 million encounters for 7.5 million patients. All 100% of the patients present age, gender, and address information, and 73% report race. Nealy 93% of patients have lab data in LOINC, 86% have medication data in RxNorm, 93% have diagnosis in SNOWMED and ICD, 81% have procedures in HCPCS or CPT, and 61% have insurance information. The richness, breadth, and diversity of this research accessible and research ready data is expected to accelerate observational studies in many diverse areas. We expect this resource to facilitate research in many years to come.

## Introduction:

The American Family Cohort (AFC) research data is managed by Stanford Center for Population Health Sciences (PHS) and converted to Observational Health Data Sciences and Informatics (OHDSI) Observational Medical Outcomes Partnership (OMOP) Common Data Model (CDM) in collaboration with Stanford Medicine Technology and Digital Solutions (TDS) team. This manuscript presents the AFC OMOP data resource description.

The AFC research data are derived from the American Board of Family Medicine PRIME Registry. The PRIME Registry, certified by Centers for Medicare and Medicaid Services (CMS) in 2016, represents over 2,500 active clinicians representing over 47 states and 8 million patients. The manuscript presents the derivation of the AFC research dataset from the PRIME registry and conversion to the OMOP data model. The final resulting dataset is hereafter referred to as AFC-OMOP.

At Stanford, our researchers now have seamless access to three large (unlinked) datasets in OMOP format: Electronic Health Record (EHR) data from Stanford adult and pediatric hospitals and their affiliates (STARR-OMOP, ~4 million patients), claims data from IBM® MarketScan® Databases (formerly known as the Truven Health MarketScan Research Databases, ~107 million patients) and EHR data from American Family Cohort (AFC-OMOP, ~8 million patients). Specific for a specialty hospital like Stanford, which sees a large number of highly complex patient cases, but has a relatively smaller pool of primary care patients, the presence of large-scale claims data and diverse EHR data provides an excellent opportunity for our researchers to validate their hypothesis and AI/ML models across a diversity of observational data assets. Furthermore, working with standardized data models such as OMOP, reduces individual researchers' burden of data pre-processing, standardization and cleaning, thereby shortening the time to insight. Furthermore, once data is converted to OMOP format, evidence can be generated using standardized analytics tools, thereby enhancing research reproducibility. Finally, a dataset such as AFC, which is collated from a large number of independent practice EHRs, benefits even more from the OMOP standardization compared to data from a single monolithic organization.

Stanford PHS is the sole provider of the AFC-OMOP research dataset. Interested researchers can reach PHS to request access (https://med.stanford.edu/phs/data.html).

# Background:

In this section, we present the background content for the AFC cohort, and Stanford's OMOP ecosystem.

## The American Family Cohort (AFC):

The American Family Cohort (AFC) is managed by Stanford Population Health Sciences (PHS). The AFC research data are derived from the American Board of Family Medicine PRIME Registry. The PRIME Registry [Philips2017], established by the American Board of Family Medicine (ABFM) and certified by CMS in 2016, is a population health and performance improvement tool for practices. It is the largest national Qualified Clinical Data Registry (QCDR) for Primary Care and was established to help smaller, independent practices meet Medicare and Medicaid reporting requirements such payment models fast-tracked changes from fee-for-service to value-based payments. PRIME pulls the data from practice EHRs and enables participating practices to select and track specific quality measures on PRIME dashboards. For example, PRIME Registry is able to extract the necessary data from practice EHR and streamline reporting for the Merit-Based Incentive Payment System (MIPS) and Primary Care First (PCF) or calculate patient risk scores based on the Charleson Comorbidity Index or Clinician Intuition scoring. Specific to population health, the PRIME dashboard allows practices to explore the social determinants of health that impact their patients and guidance on where they may find resources to address them. It allows the practice to examine each patient's care quality gaps in conjunction with Social Deprivation Index data for insight into the patient's community characteristics.

As of 2020, PRIME Registry has over 2,500 active clinicians participating from 47 States and data on approximately 8 million patients. PRIME collects data related to patient encounters found in the practice's Electronic Health/Medical Record (EHR or EMR), such as Patient Demographics (Patient Race, Ethnicity, Social History) , Medication (RxNorm Codes, NDC Codes, and CVX Codes), Diagnosis codes (ICD 9/10, SNOMED-CT, CPT codes, HCPCS), Vital Signs, Plan of Care (Cessation and counseling data), Lab Results and Lab Test Data, LOINC, Patients Notes, and Insurance related data. The PRIME registry is backed by the cloud-based FIGmd platform (https://www.figmd.com/). Founded in 2010, FIGmd software now interfaces with over 250 clinical systems, supporting over 10,000 specific clinical data elements derived from over 150,000 providers. FIGmd is accredited by Electronic Healthcare Network Accreditation Commission (EHNAC) and PRIME is hosted on FedRAMP certified AWS GovCloud. PRIME data are either securely extracted via FIGmd software or sent securely via the EHR vendor.

PRIME registry has formed a HIPAA-compliant research consortium with Stanford University that enables multiple research relationships with other universities and academic medical centers. The research consortium is committed to OMOP transformation and Safe Harbor PHI scrubbing of the OMOP data to increase data utility for research while minimizing risk to patient and practice confidentiality.

## Stanford's OMOP ecosystem strengths

In 2017, Stanford CTSA (NCATS' Clinical and Translational Science Awards program) hub, moved to adopt OMOP CDM for its second generation research Clinical Data Warehouse (r-CDW). Up until that time, Stanford Medicine had supported an in-house r-CDW, STRIDE [Lowe2009]. While STRIDE, currently branded as STARR Tools, was and continues to be hugely successful at Stanford, it was important for our CTSA to participate in broader collaborations and STRIDE's in-house data model posed barriers. In the early years, the rationale to go with OMOP was backed by successes elsewhere. For example, OMOP had demonstrated applicability for many different use cases including a) claims and EHR [Overhage2012], b) EHR based longitudinal registries [Garza2016] and, c) Hospital transactional database [Makadia2014]. Furthermore, OMOP CDM had demonstrated strong results in comparative effectiveness research [Ogunyemi2013] with minimal information loss during data transformation [Voss2015], sped up implementation of clinical phenotypes across networks [Hrispack2019], and promoted research reproducibility [Zhao2018].

Between 2017 and 2019, Technology and Digital Solutions (TDS) team undertook the journey of developing its second generation r-CDW using OMOP. During this time, it converted its adult and pediatric hospital EHR data OMOP format [Datta2020] and implemented the OHDSI ATLAS Cohort Analysis tool. The ATLAS tool was optimized for Google Cloud operations and resulted in significant performance for large cohorts containing millions of patients. Stanford TDS has developed powerful cloud-based technologies to streamline r-CDW PHI scrubbing [Datta2020], including privacy-preserving PHI scrubbing of clinical text [Datta2020]. The data from Stanford adult and pediatric hospitals and its affiliate network, ~3 million patients, is converted to OMOP and refreshed weekly. The OMOP r-CDW is referred to as STARR-OMOP.

STARR-OMOP was able to achieve one of the highest ranks in the NCATS Common metrics. In our latest report, 100% of the ~3 million patients in the database present at least one encounter having their age and date of birth on record. 64.79% of the patients have a diagnosis (ICD 9/10), over 45.49% have medication information (RxNorm), ~78.72% have lab information (LOINC and Measurements), and over 73.51% of patients have clinical notes. The development team maintains a data dictionary, technical specification, detailed Extract, Transform, Load (ETL) documentation, FAQs, and more.

Since STARR-OMOP, PHS and Stanford TDS have collaborated to transform other datasets in OMOP, specifically claims dataset such as MarketScan and Optum.

There is also a comprehensive strategy to support researchers who are using OMOP. We have up-to-date detailed documentation, FAQs, slack channel support, reusable python notebooks, training workshops, office hours and more. Since 2019, the OMOP research community has grown from 75 to over 350 researchers. Stanford OMOPs have participated in a number of network studies [Burn2020, Jin2020, Kostka2022, Morales2022, Prats-Uribe2021, Recalde2021, Reyes2021, Roel2021, Shoaibi2022, Talita2021, Tan2021] and population health and artificial intelligence research [Agarwal2022, Gao2022, Lu2022, Pfohl2021]. STARR-OMOP has also provided bedside consultation service at Stanford Health Care [Callahan2021] and is now routinely used in teaching curriculum e.g., BIOMEDIN 215: Data Science for Medicine.

The original raw data from PRIME are stored and processed in HIPAA compliant Google Cloud Platform (GCP) instance that is managed by Stanford Research Computing by a very small team of data managers who follow best practices of data provenance management, code management, testing, deployment, data lifecycling. The final OMOP datasets are Google Cloud BigQuery datasets, that are programmatically accessible to researchers on PHS Data Portal (https://redivis.com/StanfordPHS/datasets), a HIPAA compliant platform powered by Stanford University Library's data platform partner, Redivis. Secure computational environments include Stanford's HIPAA compliant research computing platforms, Nero GCP [Datta2020] and Carina (https://carinadocs.sites.stanford.edu/), the second generation of the previously presented Nero on-prem [Datta2020]. All platforms allow researchers to perform typical data science tasks using a variety of tools such as SQL, R and Python. The OHDSI tools are accessible on Carina.

# AFC-OMOP Methods:

This section will present the broad methodology used to derive the AFC dataset and then will describe the OMOP conversion and PHI scrubbing techniques.

## Generating AFC from PRIME Registry:

Since the inception of their collaboration in 2019, the American Board of Family Medicine (ABFM) and the Stanford Center for Population Health Sciences (PHS) at Stanford University have been working diligently to establish the American Family Cohort (AFC) using data from the PRIME Registry. The process involves an electronic extraction of data directly from the electronic health records (EHRs) through digital portals. This dataset comprises both structured and unstructured data, an attribute common to disparate EHRs. The data encompasses various elements including patient demographics, diagnoses, treatments, such as medications and therapies, and encounter-specific information.

The extracted data currently includes approximately 800 primary care practices spread across all 50 states, which use a variety of EHRs. Roughly half of the practices provide immediate data access, while the remainder typically have a delay of 1-2 months before the data can be received. Once collected, the data is consolidated upstream before being forwarded to Stanford's PHS. The raw data is received incrementally every two months in over 3 million individual Parquet files into a Google Cloud Platform (GCP) project. This data is imported into BigQuery (BQ) and appended to the cumulative dataset. Subsequently, the data is segmented into several versions, each bearing different levels of risk, ranging from very high risk to low risk.

Note that OMOP data model does not allow for very high risk variables such as patient names. So, such high risk data are redacted from the version that is subsequently used for downstream OMOP conversion workflow.

## Generating AFC-OMOP:

Stanford has taken the best practices of Stanford OMOP ecosystem to derive and support the AFC-OMOP. The Stanford OMOP ETL adopted the Common Workflow Language (CWL, https://www.commonwl.org/) and Cromwell workflow execution engine (https://github.com/broadinstitute/cromwell). In the years following release of STARR-OMOP, community momentum has shifted to the WDL language (https://openwdl.org/) and support for CWL has also been removed from the Cromwell workflow engine since version 85.

The AFC-OMOP implementation was written in the WDL workflow language with MiniWDL (https://github.com/chanzuckerberg/miniwdl) as the workflow execution engine. A supporting library, WDL-Kit (https://github.com/susom/wdl-kit) was created to support ETL-specific tasks (dataset creation, SQL execution, database backup, etc.) An additional feature was created to support editing SQL queries stored in YAML configuration files, as the WDL language did not support multi-line strings at the time of implementation (multi-line strings are now natively supported in the development branch of the WDL specification).

In the first release, we populate the following tables: care_site, condition_era, condition_occurrence, death, device_exposure, drug_era, drug_exposure, measurement, observation, observation_period, payer_plan_period, person, procedure_occurrence, provider, location, visit_occurrence.

The main goal in converting the AFC research data to the semantically harmonized, standardized format of the OMOP CDM was to make the dataset available for research. There are always challenges in mapping data from one model to another. This conversion displayed some of the most common and noteworthy data challenges. One of the biggest hurdles was identifying and cleaning ambiguous data. Analyses of the data included, but were not limited to the following (i) inclusion of source tables and fields to AFC-OMOP CDM (ii) removal of source values for non-events (i.e., canceled procedures) or flavors of NULL (iii) removal of impossible data (i.e., clinical events before 1900) and (iiii) identifying and de-duplicating redundant data (Table 1). Close collaboration with the AFC data owners was necessary to ensure accurate data aggregation and cleaning rules were applied. Without their involvement, incorrect assumptions could have been made which would have decreased the quality of the conversion. Another challenge with converting the data to AFC-OMOP was semantically mapping the uncoded, free text terms to OMOP standard concepts. Some important data domains contained complex, multi-faceted, uncoded, source data (i.e., family history, allergies, etc.) which needed to be semantically mapped to OMOP standard concepts by medical doctors and other ontology experts. Semantic mapping to standard concepts allows the uncoded, free text terms to be used in a codified format. The cleaning rules, transformation logic and source data statistics are documented in detailed specifications for the conceptual transformation, in SQL for the logical transformation and release notes and source statistics for a higher-level view of the AFC-OMOP.

After transformation logic was written and data were loaded into the OMOP CDM format, rigorous, post-translation, data quality checks were performed. For these tasks we employed unit testing to ensure all the required data elements and no additional data were present per the conceptual specifications; employed the Data Quality Dashboard (DQD, https://github.com/OHDSI/DataQualityDashboard) for approximately 4,000 checks on the source data in 24 different categories and contexts to ensure data were properly transformed to the OMOP CDM; and executed the Automated Characterization of Health Information at Large-Scale Longitudinal Evidence Systems (ACHILLES, https://github.com/OHDSI/Achilles) on the transformed data to provide descriptive statistics on the CDM data. When discrepancies were found, analysis was conducted, and bug fixes were performed before the data were retested to ensure compliance to the specifications and OMOP CDM conventions. These testing procedures ensure data were accurately transformed.

| Patient identifier | Medication code | Medication name | Start date | Stop date | Vocabulary |
|---|---|---|---|---|---|
| 123 | 62939844101 | Naproxen Sodium 550mg TabletTake 1 tablet(s) by mouth q12h prn for pain #60 (Sixty) tablet(s) | 2017-03-20 | 2017-04-18 | NDDF |
| 123 | 62939844101 | Naproxen Sodium | 2017-03-20 | 2017-04-18 | NDDF |
| 123 | 849431 | Naproxen Sodium | 2017-03-20 | 2017-04-18 | RxNorm |
| 123 | 62939844101 | Naproxen Sodium 550mg Tablet\|Take 1 tablet(s) by mouth q12h prn for pain\| #60 (Sixty) tablet(s) | 2017-03-20 | 2017-04-18 | NDDF |
| 123 | 62939844101 | Naproxen Sodium 550mg Tablet Take 1 tablet(s) by mouth q12h prn for pain #60 (Sixty) tablet(s) | 2017-03-20 | 2017-04-18 | NDDF |
| 123 | 849431 | Naproxen Sodium 550mg TabletTake 1 tablet(s) by mouth q12h prn for pain #60 (Sixty) tablet(s) | 2017-03-20 | 2017-04-18 | RxNorm |

Table 1. Some patients had different medication name values coupled with different medication code values on the same dates and all referred to the same drug administration.

## Scrubbing PHI from AFC-OMOP data

The raw AFC-OMOP data is maintained as a HIPAA regulated dataset. Derivative PHI scrubbed datasets are generated for specific research projects as necessary. Technology and Digital Solutions has developed and demonstrated exceptional abilities to remove/redact one or more or all PHI using safe harbor approaches. First we identify the location of PHI fields in the structured OMOP data. Then, we identify potential unstructured fields such as clinical notes, where PHI can be present. The two sets undergo two different types of PHI scrubbing. For the former, the structured data, we use standard approaches such as masking (e.g., removal of addresses), substitution or tokenization (e.g. MRNs), and jittering (e.g. dates, so patient timeline is preserved). For the unstructured data, we combined the advances in cloud computing, with a technique called "Hiding-in-Plain-Sight (HIPS)," that is detailed in our manuscript [Datta2020]. Our method is open sourced, and is referred to as TiDE. It has a suite of Natural Language Processing (NLP) and other data mining tools that recognize personally identifying patterns such as names, medical record numbers, and social security numbers. The pattern, once identified, is substituted by a similar pattern that is not an identifier — i.e., a realistic and fictitious surrogate. For example, a patient name like Jane Austen is recognized as a female name and is replaced by a fictitious female name like Mary Smith. TiDE while exceptional at finding PHI, does not guarantee removal of all PHI. It simply makes it harder to find the real leaked PHI in the presence of surrogates.

# Results:

Following Table 2 presents the high level statistics and data description for AFC-OMOP dataset at the time of writing. The data is updated periodically and these numbers will change. At this time, the AFC-OMOP doesn't have clinical text and will be brought in shortly.

| Domain | Metric | Count | Fraction |
| --- | --- | --- | --- |
| Total patients | Total unique patients in the data warehouse | 7,529,569 | |
| Total encounters | Total unique encounters in the data warehouse | 89,811,404 | |
| Age | A1 – Unique patients with an age or date of birth value | 7,529,569 | 100.00% |
| Age | A2 – Unique patients with an age < 0 | 0 | 0.00% |
| Age | A3 – Unique patients with one or more ages >120 years | 182 | 0.00% |
| Age | A6 - Unique patients with age < 18 | 760,514 | 10.10% |
| Age | A7 - Unique patients with age 18- 34 | 1,400,898 | 18.61% |
| Age | A8 - Unique patients with age 35-49 | 1,388,391 | 18.44% |
| Age | A9 - Unique patients with age 50-64 | 1,640,947 | 21.79% |
| Age | A10 - Unique patients with age 65 and above | 2,338,819 | 31.06% |
| Age | A4 - Average years age at latest visit | 46.33 | |
| Age | A5 - Standard deviation years age at latest visit | 23.47 | |
| Gender | B1 – Unique patients with an asserted (administrative) gender value Male | 3,360,269 | 44.63% |
| Gender | B2 - Unique patients with an asserted (administrative) gender value Female | 4,145,510 | 55.06% |
| Race | M1 - Unique patients reporting their race as White | 4,749,055 | 63.07% |
| Race | M2 - Unique patients reporting their race as American Indian or Alaskan Native | 27,632 | 0.37% |
| Race | M3 - Unique patients reporting their race as Black or African American | 522,980 | 6.95% |
| Race | M4 - Unique patients reporting their race as Asian | 130,480 | 1.73% |
| Race | M5 - Unique patients reporting their race as Native Hawaiian or Other Pacific Islander | 13,029 | 0.17% |
| Race | M6 - Unique patients reporting their race as Some Other Race | 42,983 | 0.57% |
| Race | M8 - Unique patients where race is unreported | 2,043,410 | 27.14% |
| Ethnicity | N1 - Unique patients reporting their ethnicity as "Hispanic or Latino" | 708,008 | 9.40% |
| Ethnicity | N2 - Unique patients reporting their ethnicity as "Not Hispanic or Latino" | 4,134,301 | 54.91% |
| Ethnicity | N3 - Unique patients where ethnicity is not reported | 2,687,260 | 35.69% |
| Social Determinants of Health | O1 - Unique patients with address | 7,529,569 | 100.00% |
| Labs - LOINC | C1 – Unique patients with a lab test coded in LOINC | 7,013,328 | 93.14% |
| Medications | D1 – Unique patients with a medication coded in RxNorm/NDC | 6,488,458 | 86.17% |
| Diagnosis – ICD9/10 | E1 – Unique patients with a diagnosis coded in ICD 9/10 | 7,037,499 | 93.46% |
| Diagnosis - SNOMED | F1 – Unique patients with a diagnosis coded in SNOMED | 7,043,757 | 93.55% |
| Procedures – ICD 9/10 | G1 – Unique patients with a procedure coded in ICD 9/10 PCS | 264,317 | 3.51% |
| Procedures – HCPCS or CPT Domain | H1 – Unique patients with a procedure coded in HCPCS or CPT | 6,101,975 | 81.04% |
| Procedures - SNOMED | I1 – Unique patients with a procedure coded in SNOMED | 2,349,894 | 31.21% |
| Observations | K1 – Unique patients with at least one vital sign coded (height, weight, BP, BMI or temp) | 6,715,735 | 89.19% |
| Behavioral | L1 – Unique patients with a smoking status coded at least once | 2,043,665 | 27.14% |

| | L2 – Unique patients with opioid use disorder coded | | |
|---|---|---|---|
| Behavioral | L2 – Unique patients with opioid use disorder coded | 50,099 | 0.67% |
| Social Determinants of Health | L3 – Unique patients with an insurance provider | 4,600,178 | 61.09% |

Table 2: AFC-OMOP statistics, for race, we use the US Census definitions (https://www.census.gov/quickfacts/fact/note/US/RHI625221)

In summary, the resource presents approximately 90 million encounters for 7.5 million patients. All 100% of the patients present age, gender, and address information, and 73% report race. Nealy 93% of patients have lab data in LOINC, 86% have medication data in RxNorm, 93% have diagnosis in SNOWMED and ICD, 81% have procedures in HCPCS or CPT, and 61% have insurance information. The richness (total, total mapped), breadth (number of states), and diversity (race, social determinants of health) of this research accessible and research ready data is expected to accelerate observational studies in many diverse areas.

## Acknowledgement:

Using CRediT taxonomy(Contributor Roles Taxonomy, https://credit.niso.org/), we present the contributing roles for our authors -

- From Technology and Digital Solutions (TDS): Somalee Datta (Supervision, Writing - original draft), Deepa Balraj (Formal Analysis, Project administration), Joe Mesterhazy (Software, Writing)
- From Odysseus Data Service, a business partner for Technology and Digital Solutions team: Melanie Philofsky (Formal Analysis, Supervision, Writing), Anna Tsvetkova (Data Curation, Software, Validation), Elena Trach (Data Curation, Software, Validation), Shravani Priya Narra (Data Curation, Software, Validation), Oleg Zhuk (Data Curation), Mary Shamkhorskaya (Data Curation), Jim Singer (Project Administration, Resources, Supervision),
- From Stanford Center for Population Health Sciences (PHS): Ayin Vala (Data Curation, Writing), Shiying Hao (Data Curation), Isabella Chu (Conceptualization, Funding Acquisition), David Rehkopf (Conceptualization, Funding Acquisition, Writing - review and editing).

Data for this project were accessed using the Stanford Center for Population Health Sciences Data Core. The PHS Data Core is supported by a National Institutes of Health National Center for Advancing Translational Science Clinical and Translational Science Award (UL1TR003142) and from Internal Stanford funding. The transformation of the AFC data into OMOP was supported through Contract #75D30122P12974 with the Centers for Disease Control and Prevention (CDC) awarded to the American Board of Family Medicine in response to Requisition 00HCPNED-2022-66549. The content is solely the responsibility of the authors and does not represent the official views of the NIH.

The OMOP conversion and data processing pipelines leverage existing Stanford Medicine STARR platform infrastructure. STARR platform is developed and operated by Stanford

Medicine Research Technology team in TDS and is made possible by Stanford School of Medicine Research Office.

We wish to acknowledge and thank the ABFM PRIME Registry participating clinicians and the American Board of Family Medicine, without whom the American Family Cohort would not be possible.